\documentclass[aps,twocolumn,pra,superscriptaddress,amsmath,showpacs,tightenlines]{revtex4}
\usepackage{epsfig,graphicx,times}
\usepackage{amstext}
\usepackage{amsmath}            
\usepackage{amssymb}            
\usepackage{graphicx}           
\usepackage{latexsym}
\usepackage{bm}
\begin{document}
\title{Transparency and amplification in a hybrid system of mechanical resonator and circuit QED}

\author{Hui Wang}
\affiliation{Institute of Microelectronics, Tsinghua University,
Beijing 100084, China}

\author{Hui-Chen Sun}
\affiliation{Institute of Microelectronics, Tsinghua University,
Beijing 100084, China}

\author{Jing Zhang}
\affiliation{Tsinghua National Laboratory for Information Science
and Technology (TNList), Tsinghua University, Beijing 100084,
China} \affiliation{Department of Automation, Tsinghua University,
Beijing 100084, China}

\author{Yu-xi Liu}
\affiliation{Institute of Microelectronics, Tsinghua University,
Beijing 100084, China} \affiliation{Tsinghua National Laboratory for
Information Science and Technology (TNList), Tsinghua University,
Beijing 100084, China}

\date{\today}

\begin{abstract}
We theoretically study the transparency and amplification of a weak
probe field applied to the cavity in hybrid systems formed by a
driven superconducting circuit QED system and a mechanical
resonator, or a driven optomechanical system and a superconducting
qubit. We find that both the mechanical resonator and the
superconducting qubit can result in the transparency to a weak probe
field in such hybrid systems when a strong driving field is applied
to the cavity. We also find that the weak probe field can be
amplified in some parameter regimes. We further study the
statistical properties of the output field via the degrees of
second-order coherence. We find that the nonclassicality of the
output field strongly depends on the system parameters. Our studies
show that one can control single-photon transmission in the
optomechanical system via a tunable artificial atom or in the
circuit QED system via a mechanical resonator.
\end{abstract}

\maketitle \pagenumbering{arabic}

\section{Introduction}
It is well known that an incident light through a cavity without
internal loss can be completely transmitted (or reflected) when it
is resonant with (or far from resonance to) the mode of the cavity
in the steady-state. However, this situation can be changed by
impurities inside the cavity or unstable boundary condition of the
cavity. When these impurities are considered as two-level atoms, a
cavity quantum electrodynamics system can be formed. The unstable
boundary condition can be described by an oscillating mirror at the
one-end of the cavity, and then the cavity together with an
oscillating mirror becomes an optomechanical system.

In optomechanical systems~\cite{Vahala2008}, when a strong driving
field is applied to the cavity, there is an analogue of
electromagnetically induced transparency (EIT) for the output at
the frequency of the weak detecting field~\cite{Agarwal2010}. Such
EIT phenomenon is equivalent to that in two coupled harmonic
oscillators~\cite{Alzar}, which has been demonstrated in
metamaterials~\cite{Tassin}. The optomechanically induced EIT and
slow light have been experimentally
demonstrated~\cite{Kippenberg2010,Teufel2011,Naeini2011}. However
when an atomic ensemble is coupled to a cavity field in the
optomechanical system, it was showed that two-level atomic
ensemble can not only be used to enhance the photon-phonon
coupling through the radiation pressure~\cite{Ian}, but also be
used to broaden the transparency windows~\cite{zhouling2011}. We
have also showed that the EIT in a three-level atomic ensemble,
which is placed inside a cavity of the optomechanical system, can
be significantly changed by the oscillating
mirror~\cite{yuechang}.

With rapid progress in the research on superconducting qubits,
scientists are now studying the quantum switch (e.g.,
Ref.~\cite{sun}) by using the mechanical resonator for information
transfer between different qubits, and also experimentalists started
to demonstrate optomechanical effect in the microwave regime. For
optomechanical systems in the microwave regime, the cavity is
usually realized by a superconducting transmission line resonator,
the mechanical resonator is realized by a suspended aluminum
membrane~\cite{Teufel}, or a beam of conducting aluminium clamped on
both ends~\cite{Regal}, or a silicon nitride
nano-structure~\cite{Hertzberg2010}. Moreover, the circuit QED,
which describes the interaction between a quantized microwave field
and superconducting qubits (called as superconducting artificial
atoms), is explored for superconducting quantum information
processing~\cite{girvin,You}. We have shown that the
electromagnetically induced transparency and absorption can be tuned
in circuit QED systems by virtue of the dressed three-level
systems~\cite{Ian2010}. In contrast to the original proposal for
EIT-induced photon blockade using atomic medium~\cite{Imamoglu1997},
the strong photon-qubit interaction in circuit QED systems  makes
the photon blockade become possible at the single-atom level, and
also the photon blockade in the microwave regime has been
experimentally demonstrated~\cite{photonblockade1,photonblockade2}.
Recent studies showed that the photon blockade can be changed into
the transparency in the circuit QED system when the probe
field~\cite{liu2012} becomes strong.

\begin{figure}
\includegraphics[bb=30 260 600 450, width=9 cm, clip]{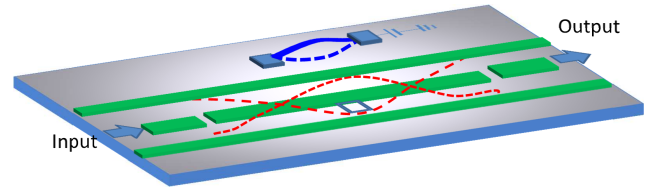}
\caption{(Color online) Schematic diagram for the circuit QED which
is coupled to a nanomechanical resonator or an optomechanical system
in microwave regime which is coupled to a superconducting qubit.}
\label{fig1}
\end{figure}

Motivated by recent studies on the nanomechanical quantum switch
between different light wavelengths using optomechanical
effects~\cite{tian,ying-dan} and progress in the circuit QED for the
strong photon-qubit coupling, we will study the photon transmission
in the circuit QED system which is coupled to a nanomechanical
resonator, or an optomechanical system which is coupled to a
superconducting qubit. Different from former
studies~\cite{Ian,zhouling2011,yuechang}, here we will focus on the
effect of single mechanical resonator (or single artificial atom) on
the EIT of the circuit QED systems (or optomechanical systems) and
statistical properties of the output field in these systems. We will
also study the amplification phenomena of the output field.

\section{Theoretical model}

\subsection{Hamiltonian}
As schematically shown in Fig.~\ref{fig1}, we study a capacitive
coupling between a nanomechanical resonator and a supderconducting
circuit QED system, which consists of a transmission line
resonator (TRL) and a superconducting qubit. Such hybrid system
can also be considered as an optomechanical system which is
coupled to a superconducting qubit. We assume that a strong
driving field with the frequency $\omega_{d}$ and a weak probe
field with the frequency $\omega_{p}$ are applied to the TLR. The
Hamiltonian of the whole system can be written as
\begin{eqnarray}\label{eq:1}
H&=&\hbar\omega_{0}c^{\dagger}c+\left[\frac{p^2}{2m}+\frac{1}{2}m\omega^{2}_{m}q^2\right]+\hbar\frac{\omega_{q}}{2}\sigma_{z}-\chi c^{\dag}cq\nonumber\\
& &+\hbar g(c^{\dag}\sigma_{-}+c\sigma_{+})+i\hbar\left(\Omega
e^{-i\omega_{d}t}c^{\dagger}
-\Omega^{\ast} e^{i\omega_{d}t}c\right)\nonumber\\
& &+i\hbar\left(\varepsilon
e^{-i\omega_{p}t}c^{\dagger}-\varepsilon^{\ast}
e^{i\omega_{p}t}c\right).
\end{eqnarray}
Here $c^{\dagger}$ and $c$ are the creation and annihilation
operators of the cavity field in the TLR with the frequency
$\omega_{0}$. $\sigma_{+}$ and $\sigma_{-}$ are the raising and
lowering operators of the superconducting qubit with the
transition frequency $\omega_{q}$, they are expressed
$\sigma_{\pm}=(\sigma_{x}\pm i\sigma_{y})/2$ by Pauli spin
operators $\sigma_{x}$ and $\sigma_{y}$. The parameters $q$ and
$p$ represent the position and momentum operators of the
mechanical resonator with the vibration frequency $\omega_{m}$ and
the mass $m$.  The coupling strengths of the cavity field to the
strong driving field and weak probe field are $\Omega$ and
$\varepsilon$, respectively. We assume $|\Omega|\gg
|\varepsilon|$. The coupling between the cavity field and the
mechanical resonator is characterized via the coupling strength
$\chi$, while $g$ denotes the interaction strength between the
superconducting qubit and the cavity field.

In the rotating reference frame at the frequency $\omega_{d}$  of
the driving field, the Hamiltonian in Eq.~(\ref{eq:1}) becomes
\begin{eqnarray}\label{eq:2}
H&=&\hbar\Delta_{1}c^{\dag}c
+\left(\frac{p^2}{2m}+\frac{1}{2}m\omega^{2}_{m}q^2\right)
+\frac{\hbar}{2} \Delta_{2}\sigma_{z}\nonumber\\
& &+\hbar g(c^{\dagger}\sigma_{-}+c\sigma_{+})-\chi c^{\dagger}cq+i\hbar(\Omega c^{\dagger}-\Omega^{\ast} c)\nonumber\\
& &+i\hbar\left[\varepsilon e^{-i\Delta
t}c^{\dag}-\varepsilon^{\ast} e^{i\Delta t}c\right]
\end{eqnarray}
with the detunings $\Delta_{1}=\omega_{0}-\omega_{d}$,
$\Delta_{2}=\omega_{q}-\omega_{d}$, and
$\Delta=\omega_{p}-\omega_{d}$.

\subsection{Langevin equations}
Based on the Hamiltonian in Eq.~(\ref{eq:2}) and phenomenologically
adding the noise and decay terms,  the Heisenberg-Langevin equations
of the variables for the mechanical resonator, the cavity field, and
the superconducting qubit can be given as
\begin{eqnarray}
\dot{q}&=&\frac{p}{m},\label{eq:3}\\
\dot{p}&=&-m\omega^{2}_{m}q-\gamma_{m}p+\chi c^{\dag}c+\xi(t),\qquad\\
\dot{c}&=&-(\gamma_{c}+i\Delta_{1})c+\frac{i\chi}{\hbar} c q-ig\sigma_{-}\nonumber\\
& &+\Omega+\varepsilon e^{-i\Delta t}+\sqrt{2\gamma_{c}}\,c_{\rm in}(t),\label{eq:5}\\
\dot{\sigma}_{-}&=&-(\gamma_{a}+i\Delta_{2})\sigma_{-}+ig\,c\,\sigma_{z}
+\sqrt{2\gamma_{a}}\,d_{\rm in}(t).\label{eq:6}
\end{eqnarray}
Here $\gamma_{c}$, $\gamma_{m}$, and $\gamma_{a}$ are the decay
rates of the cavity field, the mechanical resonator, and the
superconducting qubit. Equations of motion for the variables
$c^{\dagger}$ and $\sigma_{+}$ can be given by taking the Hermitian
conjugate of Eqs.~(\ref{eq:5}) and (\ref{eq:6}).  We do not write
down the equation of motion for the atomic operator $\sigma_{z}$
which will be explained below. $ c_{\rm in}(t)$ and $ d_{\rm in}(t)$
are the input vacuum noises of the cavity field and the
superconducting qubit, and their mean values are zeroes
\begin{eqnarray}
   \langle c_{\rm in}(t)\rangle=0,\label{eq:7}\\
  \langle d_{\rm in}(t)\rangle=0,\label{eq:8}
\end{eqnarray}
where $\langle \cdot \rangle$ represents the average over the
equilibrium state of the environment. We assume that the system
works at low temperature, and thus the eigen-energies of the qubit
and the cavity field are much higher than the thermal energy, which
means that the thermal effects on the cavity field and the qubit are
negligible. With this assumption, the vacuum noises $c_{\rm in}(t)$
and its Hermitian conjugate,  $d_{\rm in}(t)$ and its Hermitian
conjugate, satisfy the following conditions
\begin{eqnarray}
 \langle  c_{\rm in}(t) c^{\dag}_{\rm in}(t^{\prime})\rangle=\delta(t-t^{\prime}),\label{eq:9-1}\\
 \langle d_{\rm in}(t)d^{\dag}_{\rm in}(t^{\prime})\rangle=\delta(t-t^{\prime}).\label{eq:9-2}
\end{eqnarray}
Since the frequency of the mechanical resonator is much smaller
than the frequency of the cavity field and the transition
frequency of the qubit, the thermal motion of the mechanical
resonator should be taken into account. Thus we introduce the
thermal Langevin force $\xi(t)$ with the zero mean value $\langle
\xi\left( t \right) \rangle =0$  and satisfies the following
temperature-dependent correlation
function~\cite{Agarwal2010,Vitali2001}
\begin{equation}\label{eq:9-3}
\langle\xi(t)\xi(t^{\prime})\rangle=\frac{1}{2\pi}\int
\exp{\left[-i\omega(t-t^{\prime})\right]}N\left(\omega\right)
d\omega,
\end{equation}
with
\begin{equation}\label{Nomega}
N(\omega)=\hbar\gamma_{m}m\omega\left[1+\coth\left(\frac{\hbar\omega}{2k_{B}T}\right)\right].
\end{equation}

\section{Steady-state solutions and photon transmission}

We are interested in the effect of the mechanical resonator on the
photon transmission through the circuit QED system, or,
equivalently, the effect of the superconducting qubit on the
photon transmission through the optomechanical system.

\subsection{Steady-state solutions}
Using Eqs.~(\ref{eq:3}-\ref{eq:8}) and also the mean field
approximation, e.g., $\langle c q\rangle=\langle c \rangle\langle
q\rangle$, the time evolutions of the expectation values for the
operators $q$, $p$, $c$, and $\sigma_{-}$ are given by
\begin{eqnarray}
\frac{d \langle q\rangle}{ dt }&=&\frac{\langle p\rangle}{m},\label{eq:12}\\
\frac{d \langle p\rangle}{ dt }&=&-m\omega^{2}_{m}\langle q\rangle-\gamma_{m}\langle p\rangle+\chi \langle c^{\dag}\rangle\langle c\rangle,\label{eq:13}\\
\frac{d \langle c\rangle}{ dt }&=&-\left(\gamma_{c}+i\Delta_{1}-\frac{i\chi}{\hbar}\langle q\rangle\right)\langle c\rangle -i g\langle\sigma_{-}\rangle\nonumber\\
& &+\Omega+\varepsilon e^{-i\Delta t},\label{eq:14}\\
\frac{d \langle\sigma_{-}\rangle}{ dt
}&=&-(\gamma_{a}+i\Delta_{2})\langle\sigma_{-}\rangle+ig\langle c
\rangle\langle\sigma_{z}\rangle.\label{eq:15}
\end{eqnarray}
The equations of motion for the average of the operators
$c^{\dagger}$ and $\sigma_{+}$ can be obtained by taking the average
for the Hermitian conjugates of Eq.~(\ref{eq:5}) and
Eq.~(\ref{eq:6}), respectively. The nonlinear equations in
Eqs.~(\ref{eq:12}-\ref{eq:15}) cannot be solved precisely since
their steady-state solutions have infinite number of frequencies. To
approximately obtain the steady-state solutions which are exact for
the strong driving $\Omega$ and are correct to the lowest order in
the weak probe $\varepsilon$, we make the following
ansatz~\cite{book}
\begin{eqnarray}
\langle c\rangle&=&C_{0}+C_{+}e^{i\Delta t}+C_{-}e^{-i\Delta t},\label{eq:16}\\
\langle q\rangle&=&Q_{0}+Q_{+}e^{i\Delta t}+Q_{-}e^{-i\Delta t},\label{eq:17}\\
\langle \sigma_{-}\rangle&=&L_{0}+L_{+}e^{i\Delta \label{eq:18}
t}+L_{-}e^{-i\Delta t}.
\end{eqnarray}
Here $C_{\pm}$ are much smaller than $C_{0}$, and are of the same
order of $\varepsilon$. Similarly, $Q_{\pm}$ and $L_{\pm}$  are much
smaller than $Q_{0}$ and $L_{0}$. The parameters $Q_{\pm}$ and
$L_{\pm}$ are of the same order of $\varepsilon$. $C_{0}$, $Q_{0}$,
and $L_{0}$ are the steady-state solutions when $\varepsilon=0$.

By substituting Eqs.~(\ref{eq:16}-\ref{eq:18}) into
Eqs.~(\ref{eq:12}-\ref{eq:15}), and keeping the first-order terms of
$Q_{\pm}$ and $L_{\pm}$, we can give
\begin{eqnarray}
Q_{0}&=&\frac{\chi}{m\omega^{2}_{m}}|C_{0}|^{2},\label{eq:19}\\
Q_{\pm}&=&\frac{\chi}{m(\omega^{2}_{m}\pm
i\gamma_{m}\Delta-\Delta^{2})}(C_{0}C^{*}_{\mp}+C^{*}_{0}C_{\pm}),
\end{eqnarray}
by comparing the coefficients of the terms with the same frequency.
Similarly, the expressions of $L_{0}$, $L_{+}$, and $L_{-}$ can also
be obtained as
\begin{eqnarray}
L_{0}&=&\frac{i g \langle\sigma_{z}\rangle_{ss} C_{0}}{(\gamma_{a}+i\Delta_{2})},\label{eq:22}\\
L_{\pm }&=&\frac{i g \langle\sigma_{z}\rangle_{ss}
C_{\pm}}{(\gamma_{a}+i\Delta_{2}\pm i\Delta)},\label{eq:22-1}
\end{eqnarray}
where $\langle \sigma_z \rangle_{ss}$ is the steady-state value of
the operator $\langle \sigma_z \rangle$ for the superconducting
qubit.

Using Eqs.~(\ref{eq:19}-\ref{eq:22-1}), we can obtain the
steady-state values
\begin{equation}\label{eq:25}
 C_{0}=\frac{\Omega}{\gamma_{c}+i\Delta_{3}- \left[g^{2} \langle\sigma_{z}\rangle_{ss}/\left(\gamma_{a}+i\Delta_{2}\right)\right]}
\end{equation}
and
\begin{equation}\label{eq:26}
C_{-}=\frac{\varepsilon(\lambda_{1}-\lambda_{2})} {A
+2i\Delta_{3}\lambda_{3}+\lambda^{+}_{2}\lambda_{2}
-\lambda^{+}_{2}\lambda_{1}-\lambda^{+}_{1}\lambda_{2}}.
\end{equation}
The parameter $A$ in Eq.~(\ref{eq:26}) is expressed as
$A=(\lambda_{1}-\lambda_{3})(\lambda^{+}_{1}+\lambda_{3})$, and
other parameters are given as
\begin{eqnarray}
\Delta_{3}&=&\Delta_{1}-\frac{\chi^2}{m\hbar\omega^{2}_{m}}|C_{0}|^{2},\label{eq:27}\\
\lambda_{1}&\equiv&\lambda_{1}(\Delta)=\gamma_{c}-i\Delta_{3}-i\Delta+\frac{i\chi^{2}|C_{0}|^{2}}
{M(\Delta)},\label{eq:28}\\
\lambda_{2}&\equiv&\lambda_{2}(\Delta)=\frac{ g^{2}
\langle\sigma_{z}\rangle_{ss}}{(\gamma_{a}-i\Delta_{2}-i\Delta)},\label{eq:29}\\
\lambda_{3}&\equiv&\lambda_{3}(\Delta)=\frac{i\chi^{2}|C_{0}|^{2}}{M(\Delta)}.\label{eq:30}
\end{eqnarray}
with
$M(\Delta)=m\hbar(\omega^{2}_{m}-i\gamma_{m}\Delta-\Delta^{2})$.
The parameters $\lambda^{+}_{1}$ and $\lambda^{+}_{2}$ in
Eq.~(\ref{eq:26}) are defined as
\begin{eqnarray}
\lambda^{+}_{1}&\equiv&\lambda^{+}_{1}(\Delta)=[\lambda_{1}(-\Delta)]^{*},\\
\lambda^{+}_{2}&\equiv&\lambda^{+}_{2}(\Delta)=[\lambda_{2}(-\Delta)]^{*}.
\end{eqnarray}
Here we are not interested in four-wave mixing with the frequency
$\left( \omega_{p}-2\omega_{d} \right) $, thus the expression of
$C_{+}$ is not written out.

\subsection{Output field and the response of the whole system to the probe field}

The response of the system to all frequencies can be detected by the
output field, which can be given via the input-output
theory~\cite{walls,book1},
\begin{eqnarray}\label{eq:31}
c_{\rm in}\left( t \right)+c_{\rm out}\left( t \right)
=\sqrt{2\gamma_{c}} c -\frac{1}{\sqrt{2\gamma_{c}}} \left( \Omega
+\varepsilon e^{-i\Delta t} \right).
\end{eqnarray}
Using Eq.~(\ref{eq:16}) and considering the zero mean value for
the vacuum input field, i.e., $\langle c_{\rm in}(t) \rangle=0$,
we can express the mean value of the output field via
Eq.~(\ref{eq:31}) as
\begin{eqnarray}\label{eq:32}
\sqrt{2\gamma_{c}}\langle c_{\rm out}(t)\rangle
&=&(2\gamma_{c}C_{0}-\Omega)+\left(2\gamma_{c}\frac{C_{-}}{\varepsilon}-1\right)\varepsilon e^{-i\Delta t}\nonumber\\
&+&2\gamma_{c}\left(\frac{C_{+}}{\varepsilon^{*}}\right)\varepsilon^{*}
e^{i\Delta t}.
\end{eqnarray}
We can find that the second term in the right side of
Eq.~(\ref{eq:32}) corresponds to the response of the whole system to
the probe field with the frequency $\omega_{p}$. Thus the real and
imaginary parts of the amplitude of this term describe the
absorption and dispersion of the whole system to the probe field.
Since a constant does not change the lineshape of a signal, we can
define the amplitude of the rescaled output field corresponding to
the probe field $\varepsilon e^{-i\omega_{p}t}$ as
\begin{equation}\label{eq:33}
\varepsilon_{\rm out}=\frac{2}{\varepsilon}\gamma_{c}C_{-}.
\end{equation}
The real and imaginary parts of $\varepsilon_{\rm out}$ are given
as
\begin{eqnarray}
 \mu_{p}&=&\frac{\gamma_{c}(C_{-}+C^{*}_{-})}{\varepsilon},\nonumber\\
 \nu_{p}&=&\frac{\gamma_{c}(C_{-}-C^{*}_{-})}{i\varepsilon}.
\end{eqnarray}
These two quadratures of the output field can be measured by
homodyne detections~\cite{book1}. Because we are interested in the
photon transmission in the large detuning between the cavity field
and superconducting qubit, thus, without loss of generality,
hereafter we assume that the qubit always keeps at the excited
state, that is, the steady-state value
$\langle\sigma_{z}\rangle_{ss}$ is assumed as
$\langle\sigma_{z}\rangle_{ss}=1$.

\begin{figure}
\includegraphics[bb=10 150 580 695, width=4.2 cm, clip]{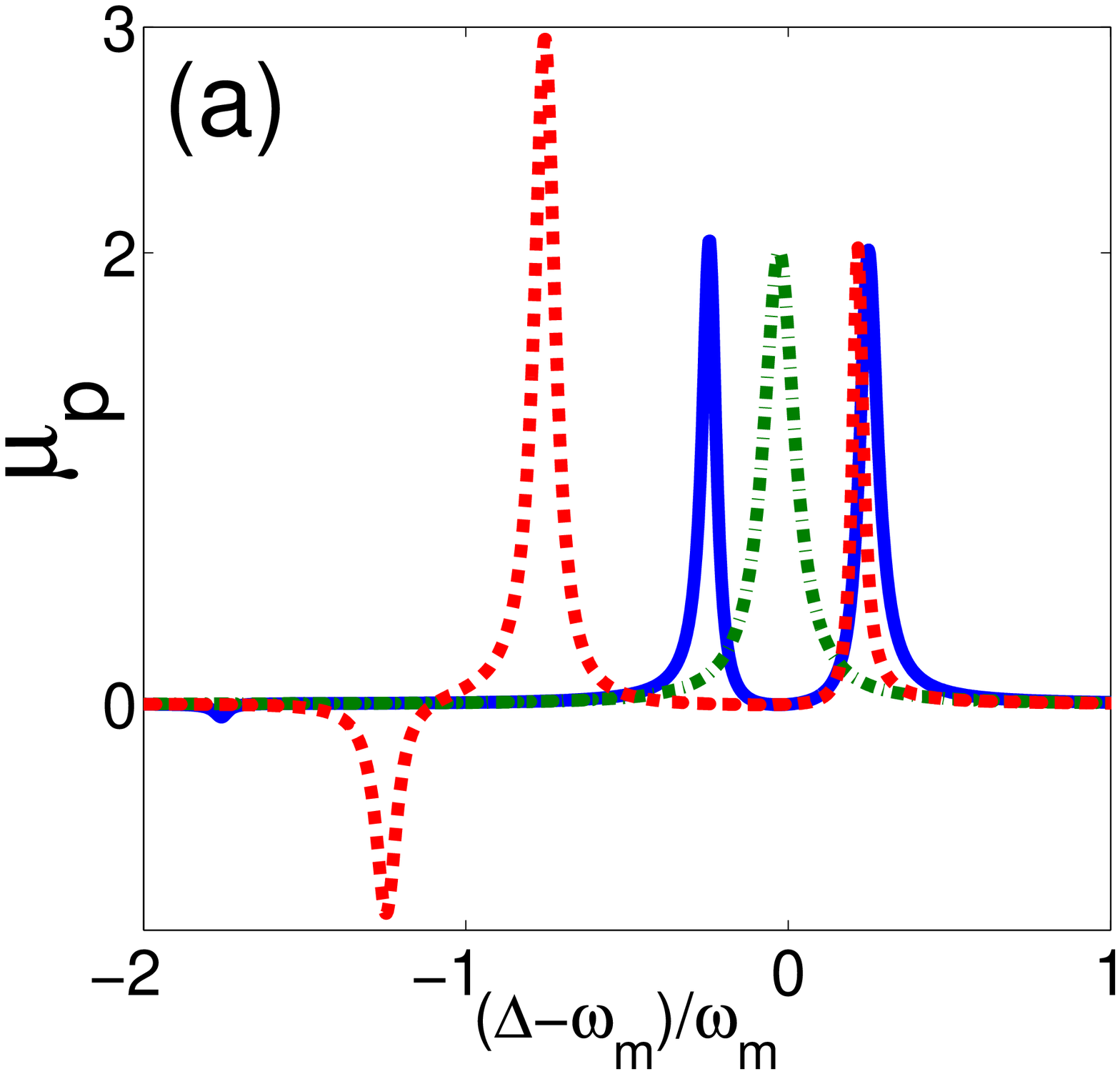}
\includegraphics[bb=32 170 580 690, width=4.3 cm, clip]{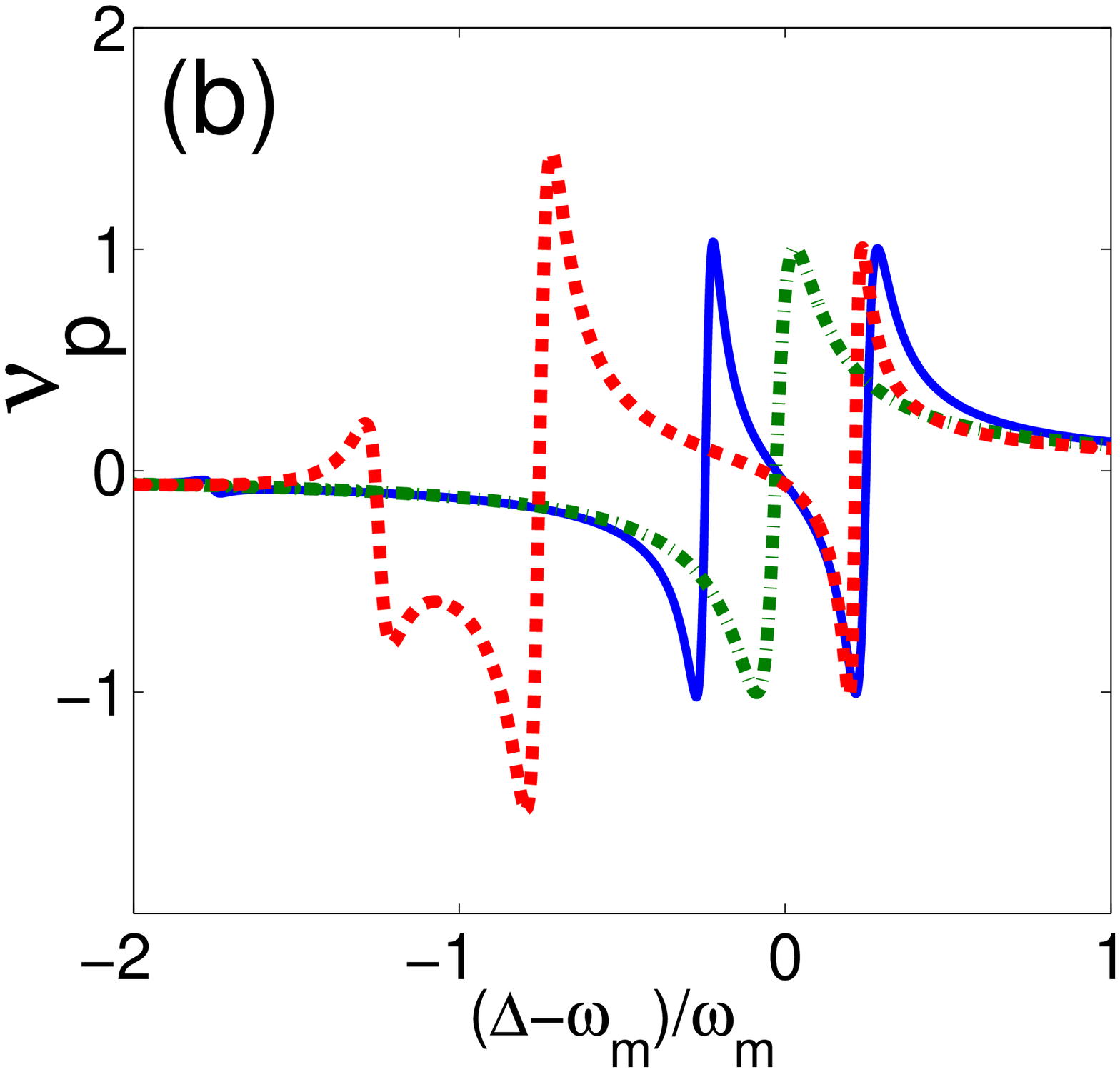}
\caption[]{(Color online) Real and imaginary parts $\mu_{p}$ and
$\nu_{p}$ of the rescaled output field $\varepsilon_{\rm out}$
given in Eq.~(\ref{eq:33}) versus the normalized detuning
parameter $\left( \Delta - \omega_m \right) / \omega_m$ are
plotted in (a) and (b) with different parameters: (i) $g=0$ and
$\chi/2\pi=2.8 \times 10^{-14}$ J/m (blue solid curve); (ii)
$g/2\pi=41.7$ MHz and $\chi=0$ (green dash-dotted curve); and
(iii) $g/2\pi=41.7$ MHz and $\chi/2\pi=2.8\times 10^{-14}$ J/m
(red dashed curve). The frequency and damping rate of the cavity
are $\omega_{0}/2\pi=5$ GHz and $\gamma_{c}/2\pi=0.5$ MHz. The
transition frequency and decoherence rate of the superconducting
qubit are $\omega_{q}/2\pi=4$ GHz and $\gamma_{a}/2\pi=1$ MHz. The
parameters of the mechanical resonator are assumed as
$\omega_{m}/2\pi=8.5$ MHz, $\gamma_{m}/2\pi=25$ Hz, and $m=2\times
10^{-15}$ kg. The frequency of the driving field is chosen to be
$\omega_{d}/2\pi=4.99$ GHz. The coupling strength between the
cavity field and the driving field is designated as $\Omega/2\pi
=3.1$ MHz. }\label{fig2}
\end{figure}

\begin{figure}
\includegraphics[bb=50 180 580 660, width=4 cm, clip]{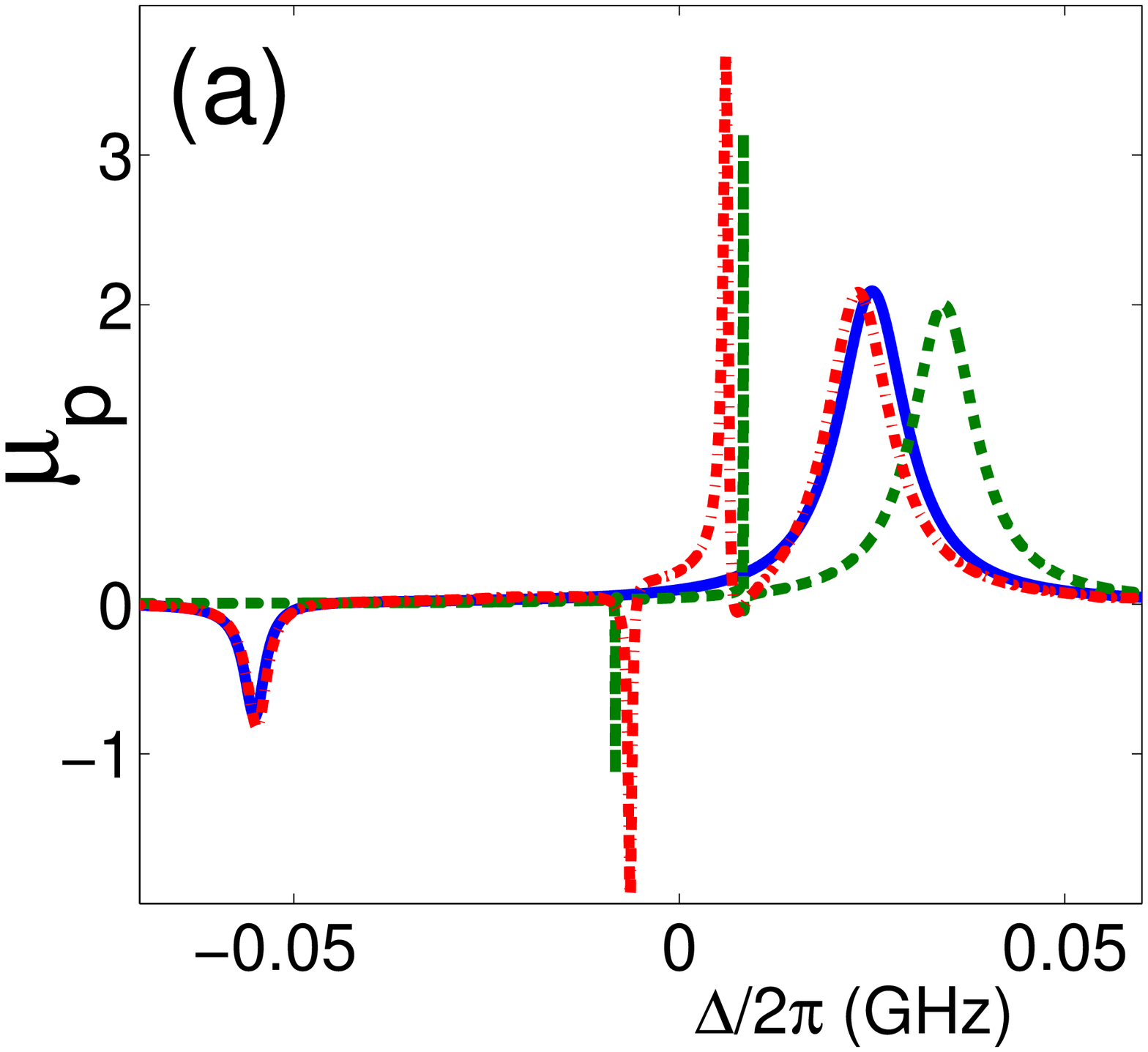}
\includegraphics[bb=60 220 580 670, width=4.2 cm, clip]{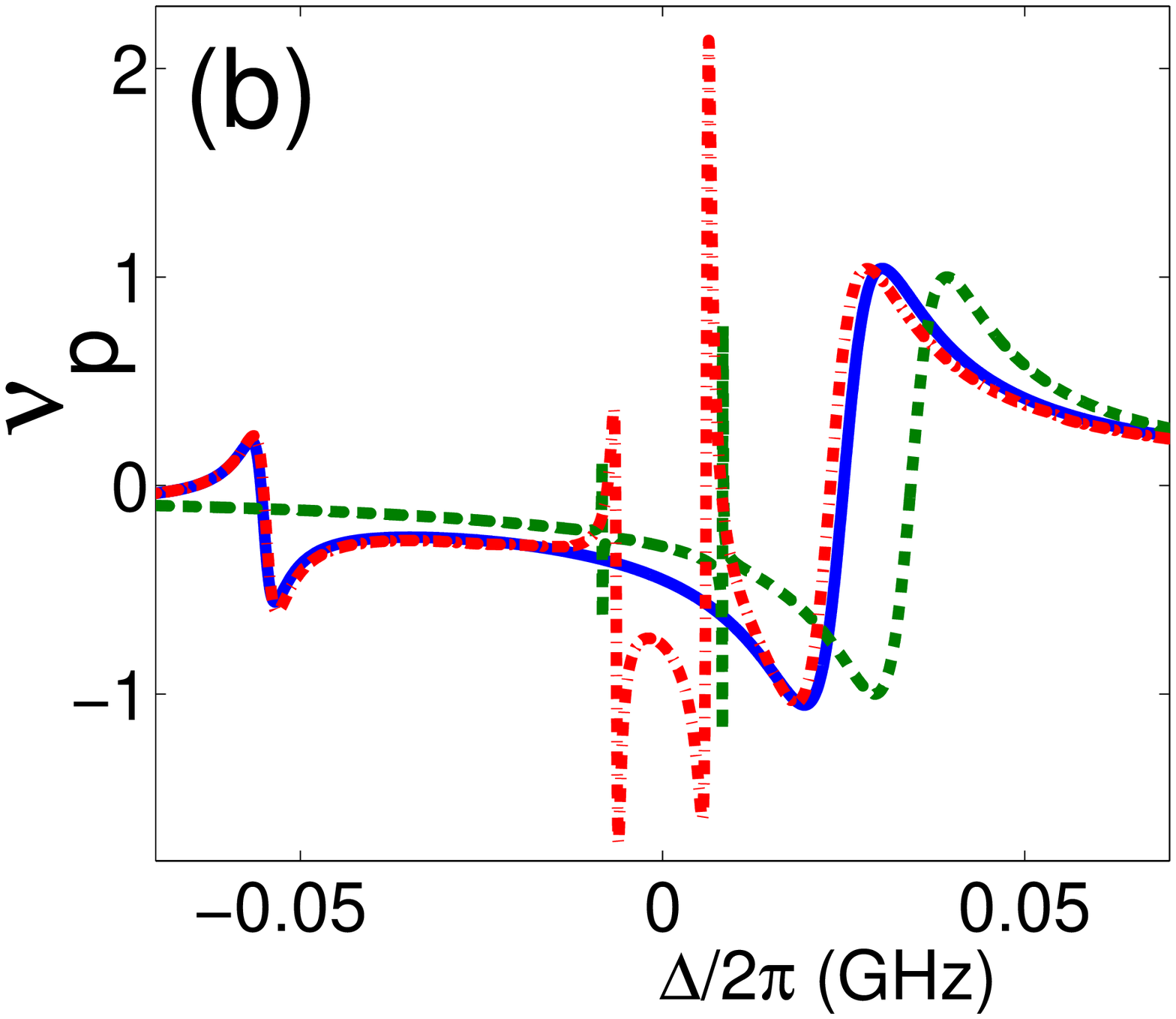}
\caption[]{(Color online) Real and imaginary parts $\mu_{p}$ and
$\nu_{p}$ of the output field $\varepsilon_{\rm out}$ given in
Eq.~(\ref{eq:33}) versus the detuing
$\Delta=\omega_{p}-\omega_{d}$ are plotted in (a) and (b) with
different parameters: (i) $g/2\pi= 30$ MHz and $\chi=0$ (blue
solid curve); (ii) $g=0 $ and $\chi/2\pi=3.0\times 10^{-13}$ J/m
(green dashed curve); and (iii) $g/2\pi=30$ MHz and
$\chi/2\pi=3\times 10^{-13}$ J/m (red dash-dotted curve). The
parameters of the cavity are assumed as $\omega_{0}/2\pi=5$ GHz
and $\gamma_{c}/2\pi=5$ MHz. The parameters of the superconducting
qubit are taken as $\omega_{q}/2\pi=4.9$ GHz and
$\gamma_{a}/2\pi=2$ MHz. The parameters of the mechanical
resonator are designated as $\omega_{m}/2\pi=8.5$ MHz,
\,$\gamma_{m}/2\pi=25$ Hz, and $m=2\times 10^{-15}$ kg. The
frequency of the driving field is assumed as
$\omega_{d}/2\pi=4.965$ GHz. The coupling strength between the
driving field and the cavity field is $ \Omega/2\pi =0.98$
MHz.}\label{fig3}.
\end{figure}

\subsection{Numerical simulations: electromagnetically induced transparency and amplification}

\begin{figure}
\includegraphics[bb=40 190 600 680, width=4.3 cm, clip]{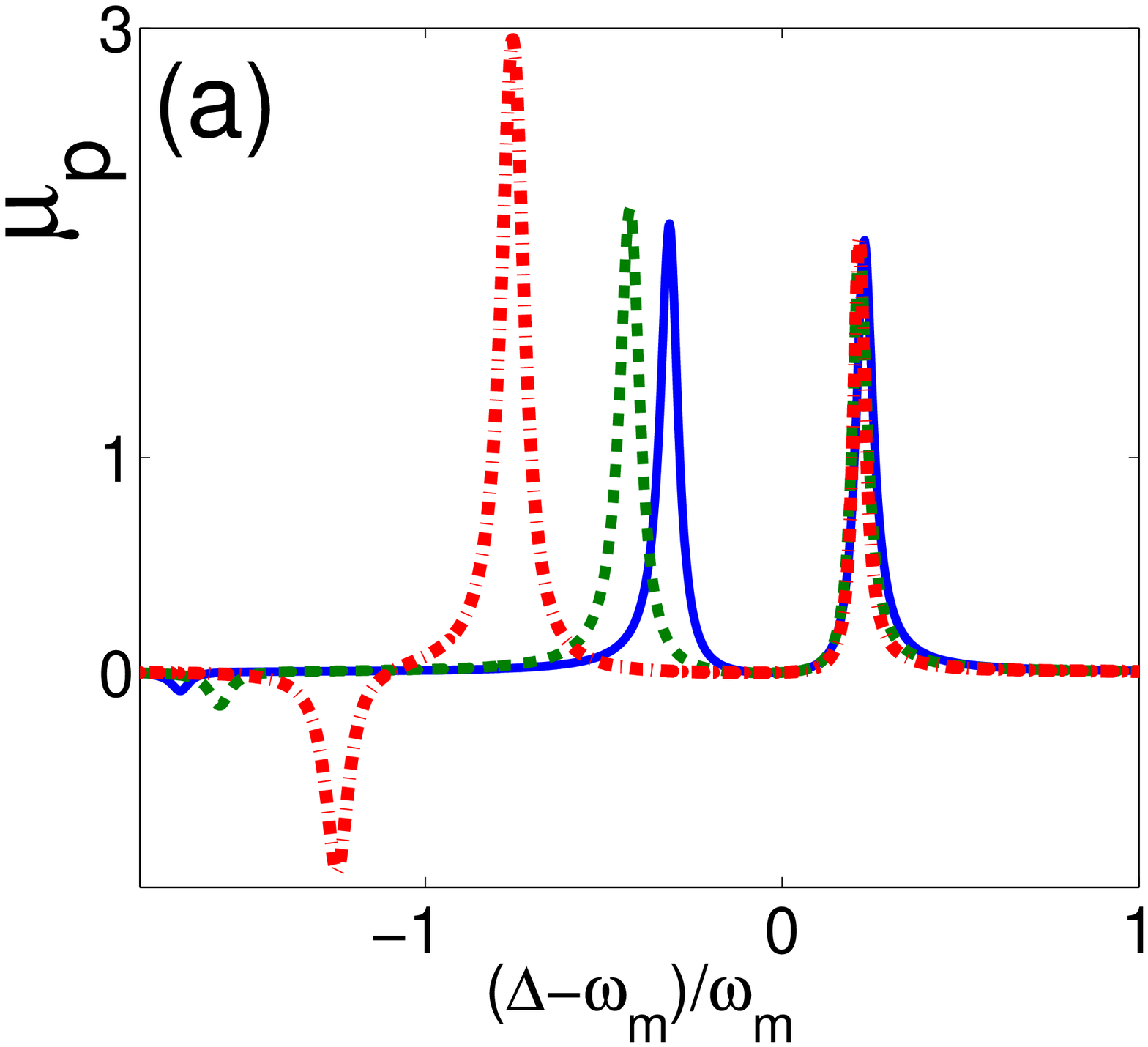}
\includegraphics[bb=15 164 570 680, width=4.1 cm, clip]{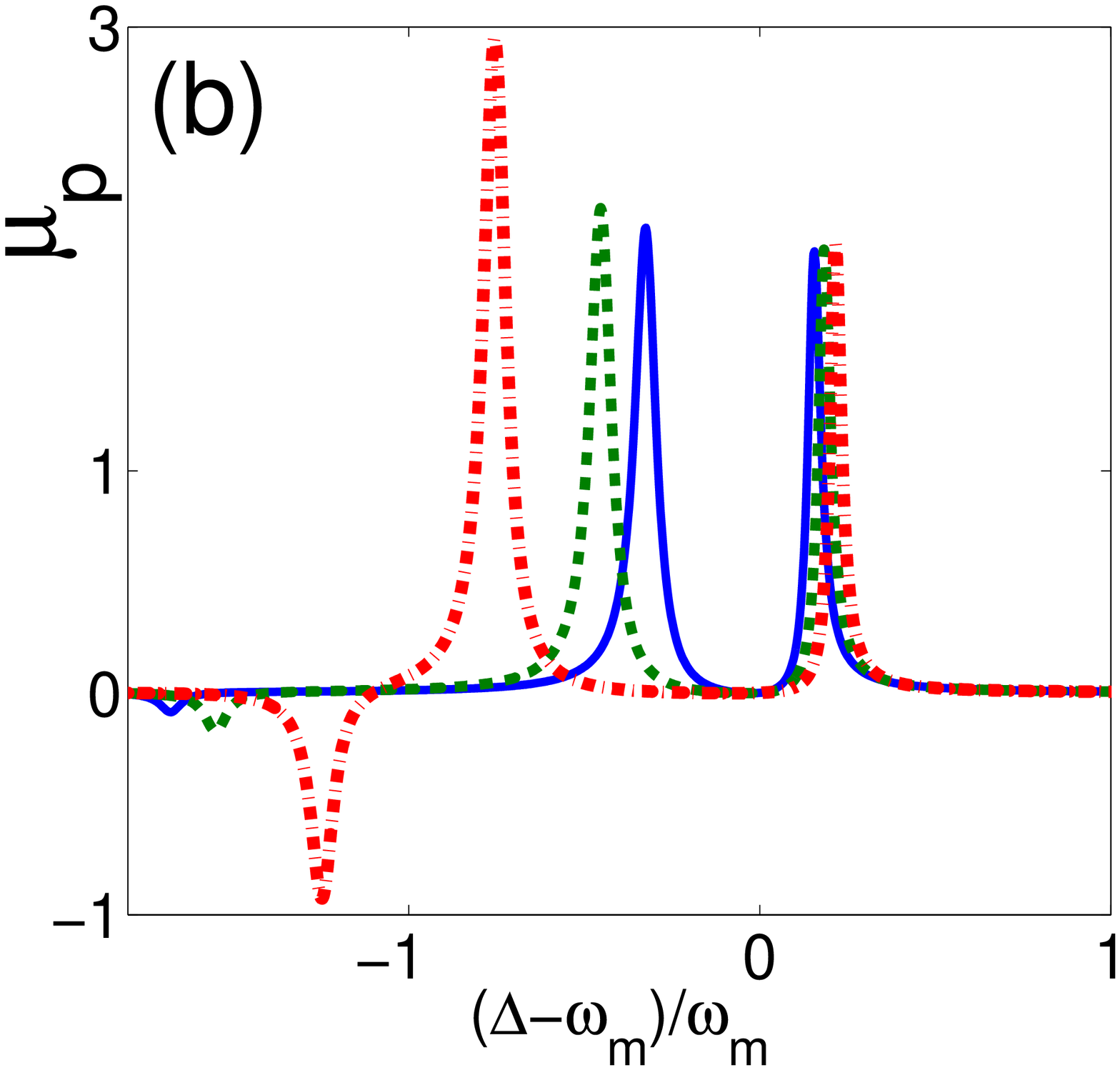}
\caption[]{(Color online) The real part $\mu_{p}$ of the output
field given in Eq.~(\ref{eq:33}) versus the normalized detuning
parameter $\left( \Delta - \omega_m \right) / \omega_m$ is plotted
in (a) with different system parameters: $g/2\pi=21.7$ MHz (blue
solid curve), $g/2\pi=31.7$ MHz (green dashed curve), and
$g/2\pi=41.7$ MHz (red dash-dotted curve), with that other
parameters are the same as in Fig.~\ref{fig2}(a). Similarly, we also
plot $\mu_{p}$ in (b) with different optomechanical coupling
strengths:  $\chi/2\pi=2 \times 10^{-14}$ J/m (blue solid curve),
$\chi/2\pi=2.4\times 10^{-14}$ J/m (green dashed curve), and
$\chi/2\pi=2.8\times 10^{-14}$ J/m (red dash-dotted curve), with
that other parameters are the same as in
Fig.~\ref{fig2}(a).}\label{fig4}
\end{figure}

We now numerically simulate $\mu_{p}$ and $\nu_{p}$ in
Figs.~(\ref{fig2}-\ref{fig4}) with experimentally accessible
parameters in optomechanical system~\cite{Hertzberg2010} and also
superconducting qubit~\cite{girvin,You}. In Fig.~\ref{fig2}, we
first study the case that the cavity field and the qubit have very
large detuning $\omega_{0}-\omega_{q}$ such that
$g^2/(\omega_{0}-\omega_{q})$ is small, but there is no coupling
between the cavity field and the mechanical resonator. In this
case, we find that there is no EIT and amplification to the weak
probe field (see green dash-dotted curve in Fig.~\ref{fig2}(a)).
Second, we study the case that there is no coupling between the
cavity field and the superconducting qubit, but there is a strong
coupling between the cavity field and mechanical resonator. We
find that there is an analogue of the EIT (see blue solid curves
of Figs.~\ref{fig2}(a) and (b)) as experimentally demonstrated in
other optomechanical
systems~\cite{Kippenberg2010,Teufel2011,Naeini2011}. Moreover, we
can also observe the amplification of the weak probe field (see
small dip of the blue solid curve of Fig.~\ref{fig2}(a)). Finally,
we study the case that $g^2/(\omega_{0}-\omega_{q})$ is small but
the coupling between the cavity field and mechanical resonator is
strong. We find that the coupling of the qubit to the cavity field
modifies the absorption and dispersive curves of the
optomechanical systems (see red dashed curves in
Figs.~\ref{fig2}(a) and (b)). Particularly, the qubit broadens the
transparency windows~\cite{zhouling2011} and also helps the
optomechanical system to significantly amplify the probe field in
some parameter regimes (see dip in red dashed curve in
Fig.~\ref{fig2}(a)).

We further numerically study the case for both the bigger
$g^2/(\omega_{0}-\omega_{q})$ and the strong optomechanical
coupling in Fig.~\ref{fig3}. We find: i) with the increase of the
optomechanical coupling strength $\chi$, the shapes of the
absorption and dispersion curves become asymmetric when there is
no coupling between the qubit and the cavity field, and also the
weak probe field is amplified greatly (see green dashed curves in
Figs.~\ref{fig3}(a) and (b)); ii) the bigger
$g^2/(\omega_{0}-\omega_{q})$ can result in the EIT~\cite{liu2012}
and amplification even without the optomechanical coupling, i.e.,
$\chi=0$ (see blue solid curves in Figs.~\ref{fig3}(a) and (b));
and iii) the strong optomechanical coupling $\chi$ can
significantly modify the transparency windows and dispersive
curves even when $g^2/(\omega_{0}-\omega_{q})$ is bigger (see red
dash-dotted curves in Figs.~\ref{fig3}(a) and (b)).

In Fig.~\ref{fig4}, we study the effects of the optomechanical
coupling strength $\chi$ and the qubit-cavity coupling strength $g$
on the transparency windows by plotting the real part $\mu_p$ of the
output field given in Eq.~(\ref{eq:33}). We first fix other
parameters and study how the transparency windows change with the
coupling strength $g$ in Fig.~\ref{fig4}(a). We find that $g$
affects not only the position and width of the transparency windows
but also the gain of the output field. In particular, we can find
that the large coupling strength $g$ leads to a high gain of the
output field. We then fix other parameters and study how the
coupling strength $\chi$ affects the transparency windows in
Fig.~\ref{fig4}(b). We find that $\chi$ has similar effects on the
transparency windows and gain of the output field as the coupling
constant $g$. Thus, Fig.~\ref{fig4} tells us that both increasing
coupling constants $\chi$ and $g$ can efficiently amplify the weak
input field in some parameter regimes.

\section{Quantum statistical properties of the output field}
For the hybrid system studied here, one would like to know the
statistical properties of the output field. To demonstrate this,
let us now study the degrees of second-order coherence for the
output field. In this case, there are vacuum input field and
strong driving field, but the weak probe field is turned off.

\subsection{Small fluctuation approximation}

We assume that the whole system reaches the steady-state and there
are small fluctuations near this steady-state. As discussed in
subsection III.A, this steady-state can be obtained by the average
of system operators given in Eqs.~(\ref{eq:12}-\ref{eq:15}) with
$\varepsilon=0$. Similar to Eqs.~(\ref{eq:16}-\ref{eq:18}), we
assume that the steady-state values of the position and momentum
of the mechanical resonator are $Q_{0}$ and $P_{0}$, and the
steady-state values of the cavity annihilation operator and the
qubit ladder operator are $C_{0}$ and $L_{0}$. With these
notations, the system operators $q$, $p$, $\sigma_-$, and $c$ can
be expressed as
\begin{eqnarray}
&& q =Q_{0}+Q,\label{eq:35}\\
&& p =P_{0}+P,\label{eq:36}\\
&& \sigma_{-}=L_{0}+\sigma,\label{eq:37}\\
&& c=C_{0}+C,\label{eq:38}\label{eq:41-2}
\end{eqnarray}
where $Q$, $P$, $\sigma$, $C$ represent the fluctuation operators of
$q$, $p$, $\sigma_-$, $c$ around the steady-state values $Q_0$,
$P_0$, $L_0$, $C_0$. From Eqs.~(\ref{eq:3}-\ref{eq:6}), we can
obtain $P_{0}=0$. Moreover, the steady-state values $Q_{0}$,
$L_{0}$, and $C_{0}$ can be calculated by Eqs.~(\ref{eq:19}),
(\ref{eq:22}), and (\ref{eq:25}), respectively. It is also clear
that the mean values of all small fluctuation operators $Q$, $P$,
$\sigma$, and $C$ are zeroes.

We now assume $\varepsilon=0$ in Eqs.~(\ref{eq:3}-\ref{eq:6}) and
substitute Eqs.(\ref{eq:35}-\ref{eq:38}) into
Eqs.~(\ref{eq:3}-\ref{eq:6}), then we can linearize
Eqs.~(\ref{eq:3}-\ref{eq:6}) up to first order of fluctuation
operators, and obtain linear dynamical equations of four
fluctuation operators as:
\begin{eqnarray}
\dot{Q}(t)&=&\frac{P(t)}{m},\label{eq:39}\\
\dot{P}(t)&=&-m\omega^{2}_{m}Q(t)-\gamma_{m}P(t)+\chi C^{*}_{0}C(t)\nonumber\\
& &+\chi C_{0} C^{\dag}(t)+\xi(t),\label{eq:40}\\
\dot{C}(t)&=&-\left(\gamma_{c}+i\Delta_{1}-\frac{i\chi}{\hbar}Q_{0}\right)C(t) +\frac{i\chi C_{0}}{\hbar}Q(t)\nonumber\\
& &-ig\sigma(t)+\sqrt{2\gamma_{c}} c_{\rm in}(t),\label{eq:41}\\
\dot{\sigma}(t)&=&-(\gamma_{a}+i\Delta_{2})\sigma(t)+ig \langle \sigma_{z} \rangle_{ss} C(t)\nonumber\\
& &+\sqrt{2\gamma_{a}}d_{\rm in}(t).\label{eq:42}\qquad
\end{eqnarray}
Here, the operator $C^{\dag}(t)$ and the steady-state value
$C^{*}_{0}$ satisfy $c^{\dag}=C^{*}_{0}+C^{\dag}(t)$. The equation
of the operator $C^{\dag}(t)$ can be given by taking the Hermitian
conjugate of Eq.~(\ref{eq:41}). Using the Fourier transform with
the definition
\begin{equation}\label{eq:44a}
f(t)=\frac{1}{2\pi}\int_{-\infty}^{\infty}f(\omega)\exp(-i\omega t)
d\omega,
\end{equation}
Eqs.~(\ref{eq:39}-\ref{eq:42}) can be given in the frequency domain
as
\begin{eqnarray}
-i\omega Q(\omega)&=&\frac{P(\omega)}{m},\label{eq:43}\\
-i\omega P(\omega)&=&-m\omega^{2}_{m}Q(\omega)-\gamma_{m}P(\omega)+\chi C^{*}_{0}C(\omega)\nonumber\\
& &+\chi C_{0} C^{\dag}(\omega)+\xi(\omega),\label{eq:44}\qquad\\
-i\omega C(\omega)&=&-(\gamma_{c}+i\Delta_{1}-\frac{i\chi }{\hbar}Q_{0})C(\omega)-ig\sigma(\omega) \nonumber\\
& &+\frac{i\chi C_{0}}{\hbar}Q(\omega)+\sqrt{2\gamma_{c}}  c_{\rm in}(\omega),\label{eq:45}\qquad\\
-i\omega \sigma(\omega)&=&-(\gamma_{a}+i\Delta_{2})\sigma(\omega)+ig\langle\sigma_{z}\rangle_{ss}C(\omega)\nonumber\\
& &+\sqrt{2\gamma_{a}}d_{\rm in}(\omega).\label{eq:46}
\end{eqnarray}
Let us now define the frequency dependent parameters
\begin{eqnarray}
E(\omega)&=&(\Lambda_{1}-\Lambda_{3})(\Lambda^{+}_{1}-\Lambda_{3})+2i\Delta_{3}\Lambda_{3}\nonumber\\
& &+\Lambda^{+}_{2}\Lambda_{2}-\Lambda^{+}_{2}\Lambda_{1}-\Lambda^{+}_{1}\Lambda_{2},\\
F(\omega)&=&\lambda_{1}-\lambda_{2},\\
R(\omega)&=&\frac{1}{m\hbar(\omega^{2}_{m}-i\gamma_{m}\omega-\omega^{2})},\\
S(\omega)&=&\frac{1}{(\gamma_{a}-i\Delta_{2}-i\omega)},\\
T(\omega)&=&\frac{1}{(\gamma_{a}+i\Delta_{2}-i\omega)}.
\end{eqnarray}
Here the parameter $\Delta_{3}$ is given in Eq.~(\ref{eq:27}). The
parameters $\Lambda_{1}$, $\Lambda_{2}$, and $\Lambda_{3}$ can be
obtained through Eqs.~(\ref{eq:28}-\ref{eq:30}) by replacing
$\Delta$ with $\omega$, respectively. Using
Eqs.~(\ref{eq:43}-\ref{eq:46}) and also the Fourier transform of the
equation of $C^{\dagger}(t)$, we can obtain the solutions of the
fluctuation operators in the frequency domain. Specifically, we have
\begin{eqnarray}\label{eq:52}
C(\omega)
&=&C_{1}(\omega) c_{\rm in}(\omega)+C_{2}(\omega) c^{\dag}_{\rm in}(\omega)+C_{3}(\omega)d_{\rm in}(\omega)\nonumber\\
& &+C_{4}(\omega) d^{\dag}_{\rm
in}(\omega)+C_{5}(\omega)\xi(\omega)£¬
\end{eqnarray}
where the parameters $C_i(\omega)$, with $i=1,2,3,4,5$, are given by
\begin{eqnarray}
C_{1}(\omega)&=&\frac{\sqrt{2\gamma_{c}}F(\omega)}{E(\omega)},\nonumber\\
C_{2}(\omega)&=&\frac{i\chi^{2}C^{2}_{0}\sqrt{2\gamma_{c}}R(\omega)}{E(\omega)},\nonumber\\
C_{3}(\omega)&=&-\frac{i g\sqrt{2\gamma_{a}}F(\omega)T(\omega)}{E(\omega)},\nonumber\\
C_{4}(\omega)&=&-\frac{\chi^{2}C^{2}_{0} g\sqrt{2\gamma_{a}}R(\omega)S(\omega)}{E(\omega)},\nonumber\\
C_{5}(\omega)&=&\frac{\chi^{3}|C_{0}|^{2} C_{0} R^{2}(\omega)+i\chi
C_{0} F(\omega)R(\omega)}{E(\omega)}.
\end{eqnarray}

\subsection{Degrees of second-order coherence for output fields}
By setting $\varepsilon=0$ in Eq.~(\ref{eq:31}) and also using
Fourier transform, we can obtain the input-output relation in the
frequency domain as
\begin{eqnarray}
2\gamma_{c}c(\omega)=\sqrt{2\gamma_{c}}[c_{\rm out}(\omega)+c_{\rm
in}(\omega)]+2\pi\Omega\delta(\omega).\label{eq:56-2}
\end{eqnarray}
Thus, using Eq.~(\ref{eq:41-2}) and  Eq.~(\ref{eq:52}), the output
field in Eq.~(\ref{eq:56-2}) can be given as
\begin{eqnarray}\label{eq:56-1}
c_{\rm out}(\omega)&=&2\pi B_{0}\delta(\omega)+G(\omega)
\end{eqnarray}
with
\begin{equation}\label{eq:57-1}
B_{0}=\sqrt{2\gamma_{c}}C_{0}-\frac{\Omega}{\sqrt{2\gamma_{c}}},
\end{equation}
and
\begin{eqnarray}\label{eq:58-1}
G(\omega)&=&B_{1}( \omega ) c_{\rm in}(\omega )+B_{2}( \omega ) c^{\dag}_{\rm in}(\omega )+B_{3}(\omega )d_{\rm in}( \omega )\nonumber\\
& &+B_{4}( \omega )d^{\dag}_{\rm in}( \omega )+B_{5}( \omega )\xi(
\omega ).
\end{eqnarray}
The parameters in Eq.~(\ref{eq:58-1}) are
\begin{eqnarray}
B_{1}(\omega)&=&\sqrt{2\gamma_{c}}C_{1}(\omega)-1,\qquad\\
B_{2}(\omega)&=&\sqrt{2\gamma_{c}}C_{2}(\omega),\qquad\\
B_{3}(\omega)&=&\sqrt{2\gamma_{c}}C_{3}(\omega),\\
B_{4}(\omega)&=&\sqrt{2\gamma_{c}}C_{4}(\omega),\qquad\\
B_{5}(\omega)&=&\sqrt{2\gamma_{c}}C_{5}(\omega).
\end{eqnarray}

To demonstrate the statistical properties of the output fields, let
us now calculate the degrees of second-order coherence, which is
defined as
\begin{eqnarray}\label{eq:59-1}
g^{(2)}(\tau)&=&\frac{\langle c^{\dag}_{\rm out}(t)c^{\dag}_{\rm
out}(t^{\prime})c_{\rm out}(t^{\prime})c_{\rm
out}(t)\rangle}{\langle c^{\dag}_{\rm out}(t)c_{\rm
out}(t)\rangle\langle c^{\dag}_{\rm out}(t^{\prime})c_{\rm
out}(t^{\prime})\rangle}\nonumber
\end{eqnarray}
with $t^{\prime}=t+\tau$. Using
Eqs.~(\ref{eq:56-1}-\ref{eq:58-1}), we can obtain $g^{\left( 2
\right)} \left( \tau \right)$ with simple calculations as

\begin{eqnarray}\label{eq:59}
g^{(2)}(\tau)
&=&\frac{2|B_{0}|^2\langle G^{\dag}(t)G(t)\rangle+2\Re[B^{*2}_{0}\langle G(t^{\prime})G(t)\rangle]}{(|B_{0}|^2+\langle G^{\dag}(t)G(t)\rangle)^2}\nonumber\\
 & &+\frac{|B_{0}|^4+2|B_{0}|^2\Re[\langle G^{\dag}(t^{\prime})G(t)\rangle]}{(|B_{0}|^2+\langle G^{\dag}(t)G(t)\rangle)^2}\nonumber\\
 & &+\frac{\langle G^{\dag}(t)G^{\dag}(t^{\prime})G(t^{\prime})G(t)\rangle}{(|B_{0}|^2+\langle
 G^{\dag}(t)G(t)\rangle)^2},
\end{eqnarray}
where $\Re[\cdot]$ represents the real part of the complex number.
Note that the correlation function $R(\tau)=\langle
G^{\dag}(t)G^{\dag}(t^{\prime})G(t^{\prime})G(t)\rangle$ can be
calculated via the Fourier transform
\begin{eqnarray}\label{eq:68-1}
&&R(\tau)=\langle G^{\dag}(t)G^{\dag}(t^{\prime})G(t^{\prime})G(t)\rangle\nonumber\\
&&=\alpha\int\!\!\!\int\!\!\!\int\!\!\!\int^{+\infty}_{-\infty}
\langle G^{\dag}(\omega_{1})
G^{\dag}(\omega_{2})G(\omega_{3})G(\omega_{4})\rangle\nonumber\\
& &\times e^{-i\omega_{1} t}e^{-i\omega_{2}
t^{\prime}}e^{-i\omega_{3} t^{\prime}}e^{-i\omega_{4} t}d\omega_{1}
d\omega_{2} d\omega_{3} d\omega_{4},
\end{eqnarray}
with the normalization coefficient $\alpha=1/(2\pi)^{4}$. From
Eqs.~(\ref{eq:9-1}-\ref{eq:9-3}) and also the definition of Fourier
transform given in Eq.~(\ref{eq:44a}), we can obtain the non-zero
correlation functions of the input noise in the frequency domain as
\begin{eqnarray}
\langle c_{\rm in}(\omega^{\prime} )  c^{\dag}_{\rm in}(\omega )\rangle&=&2\pi\delta(\omega+\omega^{\prime}),\\
\langle  d_{\rm in}(\omega^{\prime} ) d^{\dag}_{\rm in}(\omega )\rangle&=&2\pi \delta(\omega+\omega^{\prime}),\\
\langle\xi(\omega)\xi(\omega^{\prime})\rangle&=&2\pi
N(\omega)\delta(\omega+\omega^{\prime}),
\end{eqnarray}
where $N\left(\omega\right)$ is defined in Eq.~(\ref{Nomega}). Since
the stochastic force $\xi$ and vacuum inputs obey Gaussian
distributions, we can calculate higher-order correlation functions
using second-order correlation functions~\cite{Borkje}. That is,
Eq.~(\ref{eq:68-1}) becomes

\begin{eqnarray}\label{eq:72-1}
4\pi^2 R(\tau)&=&4\pi^2\langle G^{\dag}(t)G^{\dag}(t^{\prime})G(t^{\prime})G(t)\rangle\nonumber\\
&=&\int\!\!\!\int^{+\infty}_{-\infty}\!\!\!Y^{*}_{12}\left(\omega_{1}\right)Y_{12}\left(\omega_{2}\right)e^{i\left(\omega_{1}-\omega_{2}\right)\tau}
d\omega_{1} d\omega_{2}\nonumber\\
&+&\int\!\!\!\int^{+\infty}_{-\infty}\!\!\!Y_{13}(\omega_{1})Y_{13}(\omega_{2})e^{-i\left(\omega_{1}-\omega_{2}\right)
\tau} d\omega_{1} d\omega_{2}\nonumber\\
&+&\int\!\!\!\int^{+\infty}_{-\infty}\!\!\!Y_{14}(\omega_{1})Y_{14}(\omega_{2})d\omega_{1} d\omega_{2},\nonumber\\
\end{eqnarray}
where the parameters in Eq.~(\ref{eq:72-1}) are given as
\begin{eqnarray}
Y_{12}(\omega)&=&N(\omega) B_{5}(-\omega )B_{5}(\omega )+B_{2}(-\omega )B_{1}(\omega )\nonumber\\
& &+B_{4}(-\omega )B_{3}(\omega ),\\
Y_{13}(\omega)&=& N(-\omega) |B_{5}(\omega )|^{2}+|B_{2}(\omega )|^{2}+|B_{4}(\omega )|^{2},\\
Y_{14}(\omega)&=& N(-\omega) |B_{5}(\omega )|^{2}+|B_{2}(\omega
)|^{2}+|B_{4}(\omega )|^{2}.
\end{eqnarray}
Let us define
\begin{eqnarray}
y_{14}&=&\frac{1}{2\pi}\int^{+\infty}_{-\infty}Y_{14}(\omega)d\omega,\\
y_{13}&=&\frac{1}{2\pi}\int^{+\infty}_{-\infty}Y_{13}(\omega)e^{i\omega \tau}d\omega,\\
y_{12}&=&\frac{1}{2\pi}\int^{+\infty}_{-\infty}Y_{12}(\omega)e^{-i\omega
\tau}d\omega,
\end{eqnarray}
then the degrees of second-order coherence can be further
simplified to
\begin{eqnarray}\label{eq:74}
g^{(2)}(\tau)&=&\frac{|B_{0}|^4+2|B_{0}|^2y_{14}+2\Re[B^{*2}_{0}y_{12}]+2|B_{0}|^2\Re[y_{13}]}{(|B_{0}|^2+y_{14})^{2}}\nonumber\\
&
&+\frac{y^{2}_{14}+|y_{13}|^{2}+|y_{12}|^{2}}{(|B_{0}|^2+y_{14})^{2}}.
\end{eqnarray}

\begin{figure}
\includegraphics[bb=50 195 580 670, width=8.5 cm, clip]{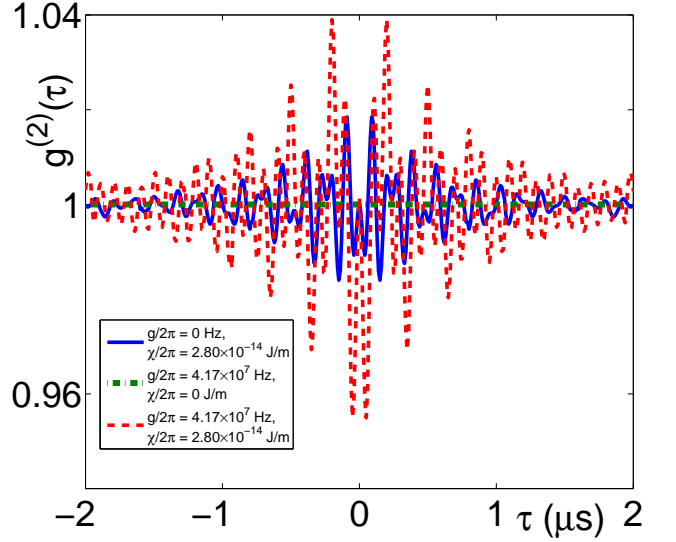}
\caption[]{(Color online) Degrees of second-order coherence
$g^{(2)}(\tau)$ are plotted as the function of the time interval
$\tau$ with different coupling constants: (i) $g=0$ and
$\chi/2\pi=2.8\times10^{-14}$ J/m (blue solid curve); (ii)
$g/2\pi=41.7$ MHz and $\chi=0$ (green dash-dotted curve); and
(iii) $g/2\pi=41.7$ MHz and $\chi/2\pi=2.8\times10^{-14}$ J/m (red
dashed curve).  Here the parameters of the cavity are taken as
$\omega_{0}/2\pi=5 $ GHz and $\gamma_{c}/2\pi=0.5$ MHz. The
parameters of the superconducting qubit are designated as
$\omega_{q}/2\pi=4$ GHz, $\gamma_{a}/2\pi=1$ MHz. The parameters
of the mechanical resonator are taken as $\omega_{m}/2\pi=8.5$
MHz, \,$\gamma_{m}/2\pi=25$ Hz, and $m=2\times 10^{-15}$ kg. The
frequency of the driving field is $\omega_{d}/2\pi=4.99$ GHz, and
the coupling strength between the driving field and the cavity
field is $\Omega/2\pi =3.1$ MHz. }\label{fig5}
\end{figure}

\subsection{Numerical simulations: Degrees of second-order coherence}

Let us numerically calculate the degrees of second-order coherence
$g^{(2)}(\tau)$ as shown in Eq.~(\ref{eq:74}) with different
coupling constants $g$ and $\chi$ in Fig.~\ref{fig5} at the zero
temperature. We find that the statistical properties of the output
field strongly depend on the parameters. For example, the output
field exhibits nonclassical properties in the case that $\chi$ is
non-zero (see, the blue solid curve in Fig.~\ref{fig5} with
$\chi/2\pi=2.8\times 10^{-14} $ J/m, and $g=0$). When $\chi=0$ and
the coupling constant $g/2\pi=41.7$ MHz, it is hard to observe
nonclassical properties of the output field. However, further
calculations of $g^{(2)}(\tau)$ show that the output field
gradually exhibits non-classical properties with the increase of
the coupling constant $g$ when $\chi=0$. This can be well
understood, because the large coupling strength $g$ results in a
large effective photon-photon interaction inside the cavity, which
leads to nonclassical properties of the cavity field as those with
the giant Kerr effects. Finally, combining the effects of two
non-zero coupling constants $\chi$ and $g$, we observe that the
coupling between the cavity field and the superconducting qubit
can enhance nonclassical properties of the output field.

\begin{figure}
\includegraphics[bb=20 200 580 680, width=8.5 cm, clip]{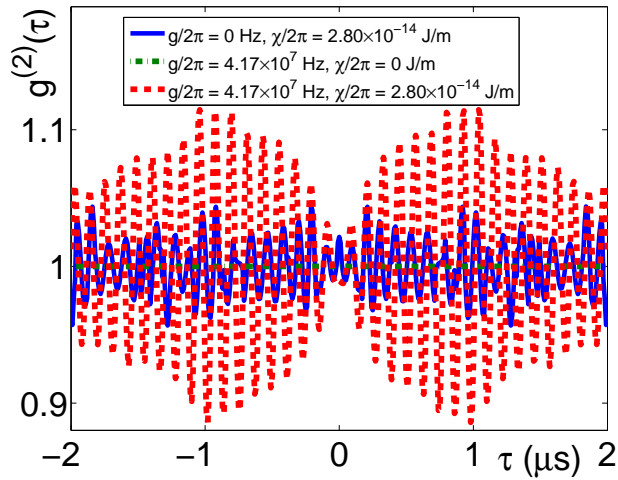}
\caption[]{Color online) Degrees of second-order coherence
$g^{(2)}(\tau)$ is plotted as a function of time interval $\tau$
at the zero temperature with different coupling constants: (i)
$g=0$ and $\chi/2\pi=2.8\times10^{-14}$ J/m (blue solid curve);
(ii) $g/2\pi=41.7$ MHz and $\chi=0$ (green dash-dotted curve); and
(iii) $g/2\pi=41.7$ MHz and $\chi/2\pi=2.8\times10^{-14}$ J/m (red
dashed curve). Here the parameters of the cavity are taken as
$\omega_{0}/2\pi=5 $ GHz and $\gamma_{c}/2\pi=0.5$ MHz. The
parameters of the superconducting qubit are designated as
$\omega_{q}/2\pi=4$ GHz, $\gamma_{a}/2\pi=1$ MHz. The parameters
of the mechanical resonator are taken as $\omega_{m}/2\pi=8.5$
MHz, \,$\gamma_{m}/2\pi=25$ Hz, and $m=2\times 10^{-15}$ kg. The
frequency of the driving field is $\omega_{d}/2\pi=4.99$ GHz, and
the coupling strength between the driving field and the cavity
field is $\Omega/2\pi =0.22$ MHz.}\label{fig6}.
\end{figure}

To study the effects of the driving field on the statistical
properties of the output field, we take all the parameters which are
the same as those in Fig.~\ref{fig5} besides the parameter $\Omega$
and plot $g^{(2)}(\tau)$ as the function of the time interval $\tau$
in Fig.~\ref{fig6} with $\Omega/2\pi=0.22$ MHz. Figure~\ref{fig6}
shows that the weak driving field enhances the nonclassical
properties of the output field. We also numerically simulate the
thermal effects for different environmental temperature $T$ of the
mechanical resonator and find that the non-classicality decreases
with the increase of the temperature $T$.

\section{Conclusions and remarks}
Our study here can be classified into two scenarios. One is a hybrid
system formed by a circuit QED system and a mechanical resonator.
Another equivalent one is the hybrid system consisting of an
optomechanical system and a superconducting qubit. We mainly focus
on the photon transmission in the case of the large detuning between
the superconducting qubit and the cavity field.

In case one, we first find out the parameters, in the circuit QED
system without the coupling to the mechanical resonator, that the
EIT can be observed as in Ref.~\cite{liu2012} in the large
detuning condition. We find that these parameters should satisfy
the strong coupling condition in the circuit QED system and
$g^2/(\omega_{0}-\omega_{q})$ should be big enough. We also find
that the weak probe field can be amplified in the output in such a
condition, while the negligibly small
$g^2/(\omega_{0}-\omega_{q})$ results in neither the transparency
nor the amplification~\cite{liu2012}. When the mechanical
resonator is coupled to the circuit QED system with the big
$g^2/(\omega_{0}-\omega_{q})$, we find that the weak coupling
between the mechanical resonator and cavity field can only
slightly changes the shape of the transparency windows of the
circuit QED system, and also slightly enhance the amplification of
the weak probe field. However, when the optomechanical coupling
becomes strong, the EIT and the amplification in the circuit QED
are strongly distorted by the mechanical resonator.

In case two,  we first find out experimentally accessible
parameters for the analogue of the EIT and amplification of the
weak probe field in optomechamical system without the coupling to
the superconducting qubit. We then study the case that the
superconducting qubit is coupled to the optomechanical system. In
this case, even with a weak coupling between the qubit and the
optomechanical system (i.e., with small but not negligible
$g^2/(\omega_{0}-\omega_{q})$), the qubit can broaden the
transparency windows, and also help the optomechanical system to
amplify the weak probe field.

We further explore the statistical properties of the output field
in different parameter regimes. We find that both (i) the strong
couplings between the cavity field and the mechanical resonator
and (ii) the big $g^2/(\omega_{0}-\omega_{q})$ can result in
nonclassical output field with the weak driving field. We also
find that both the high environmental temperature of the
mechanical resonator and strong driving field can decrease
nonclassical properties of the output field. Thus, the statistical
properties of the output field strongly depend on the properties
of each elements and the coupling strengths between different
elements in hybrid system.

Because the superconducting qubit is easier to be controlled through
external parameters and can also be coupled strongly to the cavity
field, thus in contrast to our former study~\cite{Ian} in the
optomechanical system with an uncontrollable atomic ensemble,  we
find that the properties of the photon transmission in the hybrid
system studied here can be controlled by the tunable superconducting
qubit. Our study provides a strategy for quantum switch~\cite{sun}
using a mechanical resonator (or a superconducting qubit) in an
one-dimensional chain of the coupled circuit QED systems (or the
optomechanical systems). Finally, we should also emphasize again
that this study is only specified to the large detuning between the
superconducting qubit and the cavity field, and the study for the
resonant coupling is still under way.

\section{Acknowledgement}
Y.~X.~Liu is supported by the National Natural Science Foundation of
China under Nos. 10975080 and 61025022. J. Zhang is supported by the
National Natural Science Foundation of China under Nos. 61174084 and
61134008.

\end{document}